\documentclass[runningheads]{llncs}

\usepackage[T1]{fontenc}

\usepackage{graphicx}

\usepackage{hyperref}
\usepackage{xurl}

\usepackage{tabularx} 
\usepackage{booktabs}
\usepackage{makecell}
\usepackage{multicol}
\usepackage{multirow}
\usepackage{array}
\newcolumntype{Y}{>{\centering\arraybackslash}X}
\newcolumntype{M}[1]{>{\centering\arraybackslash}m{#1}}
\usepackage{colortbl}

\usepackage{tikz}
\usepackage{pgfplots}

\pgfplotsset{
  my cycle list/.style={
    cycle list={
      {color=red,mark=*},
      {color=blue,mark=square*},
      {color=green!60!black,mark=triangle*},
      {color=orange,mark=diamond*},
      {color=purple,mark=pentagon*},
      {color=cyan,mark=o},
      {color=brown,mark=x},
      {color=magenta,mark=square},
      {color=teal,mark=triangle},
      {color=violet,mark=diamond}
    }
  }
}
\pgfplotsset{compat=1.18}
\usepackage{pgf-pie}
\definecolor{blue70}{RGB}{51,102,204}   
\definecolor{blue50}{RGB}{102,153,255}  
\definecolor{green70}{RGB}{0,153,0}     
\definecolor{green50}{RGB}{102,204,102} 
\definecolor{red70}{RGB}{204,51,51}     
\definecolor{red50}{RGB}{255,102,102}   
\definecolor{orange70}{RGB}{255,153,51} 
\definecolor{orange50}{RGB}{255,204,153}

\begin{document}

\title{Analysing Safety Risks in LLMs Fine-Tuned with Pseudo-Malicious Cyber Security Data}

\titlerunning{Analysing Safety Risks in Cyber LLMs}

\authorrunning{A.~ElZemity et al.}

\author{Adel ElZemity\orcidID{0000-0002-5402-7837} 
\and Budi Arief\orcidID{0000-0002-1830-1587} 
\and Shujun Li\orcidID{0000-0001-5628-7328}}

\institute{University of Kent, Canterbury, United Kingdom\\
\email{\{ae455, b.arief, s.j.li\}@kent.ac.uk}}

\maketitle

\begin{abstract}

Large language models (LLMs) have been used in many application domains, including cyber security. The application of LLMs in the cyber security domain presents significant opportunities, such as for enhancing threat analysis and malware detection, but it can also introduce critical risks and safety concerns, including potential personal data leakage and automated generation of new malware. Building on recent findings that fine-tuning LLMs with pseudo-malicious cyber security data significantly compromises their safety, this paper presents a comprehensive validation and extension of these safety risks using a different evaluation framework. We employ the garak red teaming framework with the OWASP Top 10 for LLM Applications to assess four open-source LLMs: Mistral 7B, Llama 3 8B, Gemma 2 9B, and DeepSeek R1 8B. Our evaluation confirms and extends previous findings, showing that fine-tuning reduces safety resilience across all tested LLMs (e.g., the failure rate of Mistral 7B against prompt injection increases from 9.1\% to 68.7\%). We further propose and evaluate a novel safety alignment approach that carefully rewords instruction-response pairs to include explicit safety precautions and ethical considerations. This work validates previous safety concerns through independent evaluation and introduces new methods for mitigating these risks, contributing towards the development of secure, trustworthy, and ethically aligned LLMs. This approach demonstrates that it is possible to maintain or even improve model safety while preserving technical utility, offering a practical path towards developing safer fine-tuning methodologies. 


\end{abstract}

\section{Introduction}

The increasing use of large language models (LLMs) in cyber security applications necessitates a rigorous examination of their benefits and potential safety risks. LLMs have shown exceptional capabilities in many text generation tasks, including code synthesis~\cite{sagodi2024methodology}, software vulnerability detection~\cite{ccetin2024empirical,ozturk2023new} and question answering~\cite{Raiaan2024A}, signalling their transformative potential across various tasks. However, this promise is accompanied by substantial safety risks, requiring focused attention from researchers and practitioners alike~\cite{charan2023text,Derner2023A,yao2024survey}.

A crucial factor in the success and utility of LLMs is their ability to maintain safety while being fine-tuned for specific domains to enhance their domain specific knowledge. While fine-tuning can enhance performance on specialised tasks, it may also introduce new vulnerabilities or amplify existing ones. This is particularly critical in cyber security applications, where the consequences of model vulnerabilities can be severe. 

Recent studies have shown how malicious actors can exploit fine-tuned LLMs to generate phishing campaigns, malware code, and other harmful content~\cite{Alotaibi2024Cyberattacks,Falade2023DecodingTT,Firdhous2023,Roy2023Generating}. Furthermore, the increasing misuse of generative AI tools like FraudGPT~\cite{Falade2023DecodingTT} and WormGPT~\cite{Firdhous2023} in cyberattacks highlights the urgent need for systematic safety analysis of fine-tuned LLMs. These tools enable adversaries to execute more sophisticated and scalable attacks, demonstrating how fine-tuning can be weaponised for malicious purposes. For instance, a recent study by Falade~\cite{Falade2023DecodingTT} revealed how malicious LLMs can be exploited to generate phishing lures, impersonation schemes and deepfakes, amplifying the arsenal of cybercriminals and exposing significant vulnerabilities.

This paper builds upon recent findings from the CyberLLMInstruct study~\cite{CyberLLMInstruct}, which demonstrated that fine-tuning LLMs with pseudo-malicious cyber security data significantly compromises their safety. While that work employed the DeepEval framework~\cite{DeepEvalDocumentation} for evaluation, we present a comprehensive validation and extension of these findings using a completely different evaluation approach. In this paper, we employ the \emph{garak} red teaming framework~\cite{garak} with the OWASP Top 10 for LLM Applications~\cite{owasp2025} (see Appendix~\ref{app:owasp} for details) to assess how fine-tuning affects model susceptibility to various vulnerabilities. Note that our evaluation covers seven out of ten OWASP vulnerability categories, as the garak framework did not yet support Supply Chain, System Prompt Leakage, and Unbounded Consumption at the time we conducted this work. 

Our analysis confirms and extends the critical safety concerns identified in deploying fine-tuned LLMs in cyber security contexts. We validate our findings using the same CyberLLMInstruct dataset~\cite{CyberLLMInstruct}, which contains 54,928 pairs of instructions and responses of pseudo-malicious cyber security data. The CyberLLMInstruct dataset is publicly available at \url{https://github.com/Adelsamir01/CyberLLMInstruct}.

The term ``pseudo-malicious'' refers to data that contains instructions and descriptions of malicious cyber security actions, but without actual harmful code. Instead, it includes step-by-step descriptions and pseudo-code of how to perform these actions, such as malware creation, social engineering techniques, and various attack methodologies. This approach allows for comprehensive security testing while maintaining ethical boundaries. 

The dataset's composition reflects real-world cyber threats, with malware-related content (35\%), social engineering and phishing (25\%), DoS/DDoS attacks (10\%), MITM attacks (10\%), zero-day exploits (8\%), password attacks (6\%), and emerging threats like IoT and injection attacks (3\% each). This distribution ensures our evaluation covers the most prevalent and critical cyber security threats while maintaining a balanced representation of different attack vectors.

We make the following \textbf{contributions} in this work:
\begin{itemize}
\item We provide independent validation of safety risks in fine-tuned LLMs using the garak red teaming framework with OWASP Top 10 for LLM Applications, confirming and extending previous findings from the CyberLLMInstruct study~\cite{CyberLLMInstruct} with a completely different evaluation methodology.

\item We demonstrate that fine-tuning on pseudo-malicious data reduces safety resilience across \emph{all} tested LLMs, including the reasoning-capable DeepSeek R1 8B model, which was not previously evaluated in this context.

\item We propose and evaluate a novel safety alignment approach that carefully rewords instruction-response pairs to include explicit safety precautions and ethical considerations, demonstrating significant improvements in model safety while preserving technical utility.
\end{itemize}

Overall, this work establishes a foundation for understanding the safety implications of fine-tuning LLMs for cyber security applications, while providing insights into safety alignment and a novel approach for improving model safety.

The rest of this paper is organised as follows. Section~\ref{sec:related} provides an overview of related work on LLM safety and recent work in safety-aware LLM fine-tuning. Section~\ref{sec:threat-model} outlines the threat model that motivates our research. Section~\ref{sec:methodology} describes our systematic methodology for evaluating safety risks in fine-tuned LLMs for cyber security applications and our novel approach to improve safety alignment. Section~\ref{sec:results} presents our results, along with a detailed analysis of the key findings and evaluations done to validate our work. Section~\ref{sec:discussions} discusses the implications of our findings, future work, and the limitations of current approaches. 
Finally, Section~\ref{sec:conclusion} concludes our paper.

\section{Related Work} \label{sec:related}

Recent research has highlighted the critical safety risks associated with fine-tuning LLMs. Several studies have investigated different aspects of this problem and proposed various mitigation strategies. 

Eiras et al.~\cite{eiras2024mimicking} demonstrated how fine-tuning can compromise safety alignment in closed LLMs, though their proposed ``Paraphrase'' mitigation strategy was found to have limitations in terms of controllability and stability. The work also raised concerns about the generalisability of mitigation approaches when the prompting strategy is unknown in advance.

Bianchi et al.~\cite{bianchi2024safetytuned} explored the trade-off between helpfulness and harmlessness in safety-tuned LLMs, documenting important observations about the safety-helpfulness tension. However, their work was limited by a relatively small safety dataset and remained susceptible to adversarial attacks. The study highlighted the need for more systematic approaches to resolve the fundamental challenge of maintaining safety while preserving model capabilities.

In an attempt to address these challenges, Zhu et al.~\cite{zhu2024lockingfinetunedllmssafety} proposed a method to locate safety vectors for fine-tuned LLMs. While their approach is promising, it was limited to proprietary API-based models and focused primarily on attention heads and the final layer, missing opportunities to explore more comprehensive safety mechanisms in intermediate layers and feed-forward networks.

More recently, Hsu et al.~\cite{hsu2024safe} introduced Safe LoRA, a method aimed at reducing safety risks during fine-tuning by projecting weights to a safety subspace. However, their approach lacked theoretical justification for the projection mechanism and was primarily evaluated on Llama models, raising questions about its generalisability to other architectures like Mistral, Phi, and Gemma. The work also used artificially augmented harmful samples rather than standard safety benchmarks, limiting its practical applicability.

These studies collectively highlight the ongoing challenges in maintaining the safety of LLM during fine-tuning, particularly in cyber security contexts where the risks are amplified. While various approaches have been proposed, significant gaps remain in understanding how different fine-tuning methods might affect model vulnerabilities and how to mitigate these risks effectively while preserving model capabilities.

Other recent work has specifically focused on safety-aware fine-tuning approaches. Choi et al.~\cite{choi2024safetyaware} proposed the SAFT framework that automatically filters harmful data during fine-tuning using matrix factorisation, but their approach was limited by its reliance on lexical overlap metrics (BLEURT and ROUGE-L) for measuring helpfulness, which may not capture the nuanced requirements of cyber security applications. 

Qi et al.~\cite{qi2024finetuning} demonstrated that safety alignment can be compromised through fine-tuning, even with benign data, but their analysis focused on general harmfulness without specific consideration of cyber security threats. 

Peng et al.~\cite{peng2024navigating} introduced the concept of ``safety landscape'' and the VISAGE metric to measure fine-tuning risks, but their evaluation primarily relied on refusal keyword detection, which may not be sufficient for complex cyber security scenarios where safety does not always mean refusing to answer. 

Jain et al.~\cite{jain2024what} provided a mechanistic study of safety fine-tuning using synthetic data, but their analysis was limited in its application to real-world cyber security datasets.

Our work addresses these limitations by: (1) using comprehensive safety metrics beyond lexical overlap, including domain-specific cyber security evaluations; (2) focusing specifically on cyber security threats and their unique safety requirements; (3) developing a more nuanced safety alignment approach that goes beyond simple refusal detection; and (4) validating our approach on a large-scale real-world cyber security dataset.

\begin{figure}[tb!]
\centering
\includegraphics[width=\linewidth]{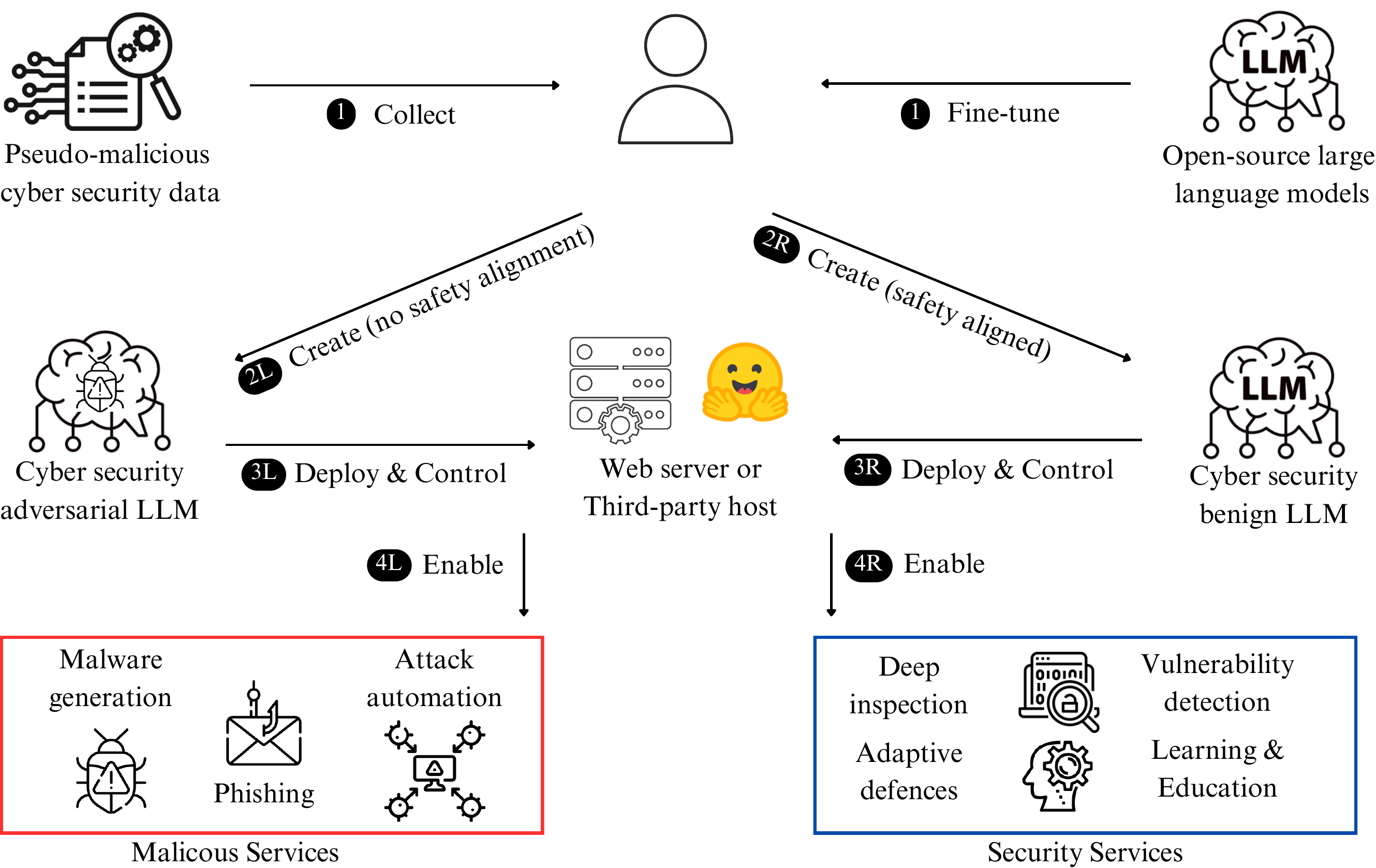}
\caption{Dual-pathway threat model showing how identical resources (pseudo-malicious cyber security data and open-source LLMs) can lead to either malicious or benign outcomes depending on safety alignment during fine-tuning.}
\label{fig:threat_model}
\end{figure}
    
\section{Threat Model}
\label{sec:threat-model}

To understand the security implications of fine-tuning LLMs with cyber security data, we develop a threat model that illustrates the dual-use nature of these technologies. This model serves as the foundation for our subsequent safety analysis and demonstrates why systematic security evaluation is crucial in this domain.

As shown in Fig.~\ref{fig:threat_model}, our threat model illustrates a comprehensive 4-step process that demonstrates how identical foundational resources can lead to fundamentally different outcomes depending on the implementation of safety alignment during model development. The model presents a dual-pathway framework centred around a person or an entity who has access to the same foundational resources: pseudo-malicious cyber security data and open-source LLMs.
The model demonstrates a structured progression through four steps:

\textbf{Step 1 -- Resource Collection and Model Preparation}: The process begins with two parallel activities: collecting pseudo-malicious cyber security data and fine-tuning open-source LLMs.

\textbf{Step 2 -- Critical Divergence Point}: This represents the most crucial decision point in the process. That person can choose between two approaches: creating models without safety alignment (Step 2L) or creating models with proper safety alignment (Step 2R). This step determines whether the development process will lead to malicious or benign outcomes.

\textbf{Step 3 -- Deployment and Control}: Both pathways proceed to deployment on web servers or third-party hosting platforms (Steps 3L and 3R). However, the control mechanisms and intended purposes differ significantly based on the safety alignment decisions made in Step 2.

\textbf{Step 4 -- Service Enablement}: The final step (Steps 4L and 4R) enables the actual services that these systems will provide, leading to two distinct categories of outcomes:

\begin{itemize}
\item \textbf{Malicious Services (Left Pathway)}: When Step 2L is chosen (no safety alignment), the system enables malicious services including malware generation, phishing campaigns, and attack automation. These capabilities can be exploited by malicious actors to conduct sophisticated cyber attacks.

\item \textbf{Security Services (Right Pathway)}: When Step 2B is chosen (safety alignment), the identical foundational resources produce legitimate security services including deep inspection, vulnerability detection, adaptive defences, and learning \& education capabilities. These systems serve defensive purposes and are utilised by security researchers and practitioners.
\end{itemize}

The central insight of this model is that the difference between malicious and benign outcomes lies not in the availability of resources—which are identical in both pathways—but in the critical decision point at Step 2 regarding safety alignment implementation. This highlights the paramount importance of safety-aware approaches in determining whether LLM development contributes to cyber threats or cyber defence.

Given the critical role of safety alignment highlighted in our threat model, it becomes essential to develop systematic methods for evaluating how different fine-tuning approaches affect model vulnerability to security risks. To assess these security risks systematically, we employ the OWASP Top 10 for LLM Applications framework~\cite{owasp2025} (see Appendix~\ref{app:owasp} for details) to evaluate how fine-tuning affects each LLM's susceptibility to various vulnerabilities. This framework, developed by experts in AI and cyber security, helps developers and organisations mitigate vulnerabilities that could lead to security breaches, data leakage, or operational failures in real-world deployments.

\section{Methodology}
\label{sec:methodology}

This section presents our approach to validating and extending the safety risk findings from the CyberLLMInstruct study~\cite{CyberLLMInstruct}. We begin by detailing our model selection and fine-tuning process, followed by safety analysis using NVIDIA’s open-source red teaming framework called \emph{garak}~\cite{garak}, against the OWASP Top 10 for LLM Applications~\cite{owasp2025}. This represents a different evaluation methodology from the DeepEval framework~\cite{DeepEvalGitHub} used in the original study. Finally, we describe our novel safety alignment approach to mitigate identified vulnerabilities.

\subsection{Model Selection and Fine-tuning}

We selected four open-source models spanning different architectures and sizes: Mistral 7B, Llama 3 8B, Gemma 2 9B, and DeepSeek R1 8B (using the DeepSeek-R1-0528-Qwen3-8B variant due to computational constraints). These models were chosen based on the work done in~\cite{CyberLLMInstruct} to verify the results with a different evaluation framework and one additional model (DeepSeek R1 8B). The range of models allowed us to investigate how architectural differences may affect both security resilience during fine-tuning.

All models were fine-tuned on the CyberLLMInstruct dataset using standard supervised fine-tuning practices. The complete technical setup, including hardware specifications, software libraries, and hyperparameter configurations, is detailed in Appendix~\ref{app:technical-setup}.

\subsection{Safety Analysis}

Building on the threat model detailed in Section~\ref{sec:threat-model}, this section describes our systematic approach to evaluating these risks using the OWASP Top 10 for LLM Applications framework.

We evaluated model safety using the garak framework~\cite{garak}. Unlike the DeepEval framework~\cite{DeepEvalDocumentation} used in the CyberLLMInstruct study~\cite{CyberLLMInstruct}, which focused mainly on refusal and harmfulness metrics, garak executes fixed, vulnerability-specific probe suites mapped to the OWASP Top 10 for LLM Applications. This enables reproducible, security-oriented testing across categories such as prompt injection, data poisoning, and sensitive information disclosure. Complete technical specifications and probe configurations are provided in Appendix~\ref{app:eval}.

\subsection{Safety Alignment}

Our results re-confirmed the results in~\cite{CyberLLMInstruct} that fine-tuning on pseudo-malicious data can significantly compromise model safety. To address this challenge, we developed a novel safety alignment approach inspired by several key past studies in LLM alignment research. Our method builds on an insight from Sun et al.~\cite{sun2024evaluating} that rewording instructions significantly affects model performance and alignment, as well as the concept of leveraging mistakes as learning opportunities reported by Chen et al.~\cite{chen2024gaining}.

The safety-regulating process involved carefully rewording each instruction-response pair in the CyberLLMInstruct dataset to incorporate explicit safety precautions and risk explanations while preserving the technical content. Specifically, each transformed entry included the following three components:
\begin{itemize}
\item explicit warnings about potential misuse and ethical implications,

\item clear statements about legal boundaries and responsible disclosure, and

\item educational context explaining defensive applications of the information.
\end{itemize}

To perform the safety-regulation at scale, we conducted a comparative analysis of several state-of-the-art LLMs. Due to the pseudo-malicious nature of CyberLLMInstruct, many commercial LLMs consistently refused to process the safety-regulating requests, citing safety concerns.

After extensive testing, we selected DeepSeek-R1~\cite{deepseekai2025} for the safety-regulating task. We initially tested models such as GPT-4o, Claude 3, and Llama-3 70B, but these either refused to process the pseudo-malicious content or were impractical to deploy at scale. This decision was driven by two key factors: first, as an open-source model, it could be deployed locally, ensuring that sensitive copyrighted information remained within our secure environment without sharing with third-party entities; second, recent studies have highlighted that DeepSeek-R1 has significantly fewer safeguards compared to other LLMs. Specifically, Arrieta et al.~\cite{arrieta2025o3minivsdeepseekr1safer} demonstrated that DeepSeek-R1 produces approximately 12\% more unsafe responses than OpenAI's o3-mini model when subjected to systematic safety testing, making it more amenable to processing our dataset while still maintaining the ability to incorporate safety elements. The safety-regulating process was manually verified for consistency and completeness.

Our approach is conceptually similar to the work by Chen et al.~\cite{chen2025secaligndefendingpromptinjection}, who demonstrated that fine-tuning on carefully reworded instruction-response pairs can dramatically improve model resilience against adversarial inputs while maintaining utility. However, to the best of our knowledge, our approach has not been previously implemented and tested on cyber security pseudo-malicious data, presenting a novel opportunity to study its effects on safety improvements in this high-risk domain. After transforming the CyberLLMInstruct dataset, we fine-tuned the same four models we chosen for the safety analysis task using the safety-aware version and evaluated the resulting models using the garak framework aligning with OWASP Top 10 for LLM Applications. We refer to these as the ``safety-enhanced models'': the same base checkpoints fine-tuned on our in-house safety-regulated CyberLLMInstruct (see Section~\ref{subsec:safety_alignment}). 

The garak evaluation measures failure rates, which are calculated as the percentage of test cases where the model produces inappropriate or harmful outputs when tested against adversarial prompts (see Appendix~\ref{app:prompts} for a link to some examples). The failure rate measures unsuccessful defences, with higher failure rates indicating greater vulnerability.

\begin{table*}[t!]
\centering
\caption{A summary of the garak failure rates of base (green), fine-tuned (red), and safety-enhanced (blue) LLMs across the seven OWASP vulnerabilities. The scores range from 0 (fully secure) to 100 (completely vulnerable). In each cell, the short vertical bar to the right of the bar chart indicates the maximum score of 100, which helps shows where the current scores are.}
\label{tab:owasp_results}
\begingroup
\tikzset{every picture/.style={scale=1.25, transform shape}}
\resizebox{\linewidth}{!}{  
\begin{tabularx}{1.1\textwidth}{@{}l*{4}{Y}@{}}  
\toprule
{\small Vulnerability} & {\small Mistral 7B} & {\small Llama 3 8B} & {\small Gemma 2 9B} & {\small Deepseek R1 8B}\\
\midrule
\multicolumn{1}{@{}l}{\makecell[bl]{\vspace{-0.03cm}\small Prompt\\\vspace{-0.02cm}Injection}} & 
  \begin{tikzpicture}[baseline=(current bounding box.south)] 
      \draw[fill=green!60] (0,0.18) rectangle (0.091,0.32); 
      \node[anchor=east, scale=0.6] at (-0.15,0.25) {9.1};
      \draw[fill=red!60] (0,-0.02) rectangle (0.687,0.14);
      \node[anchor=east, scale=0.6] at (-0.15,0.06) {68.7};
      \draw[fill=blue!60] (0,-0.18) rectangle (0.063,-0.06);
      \node[anchor=east, scale=0.6] at (-0.15,-0.12) {6.3};
      \draw (1,0.32) -- (1,-0.18);
  \end{tikzpicture} &
  \begin{tikzpicture}[baseline=(current bounding box.south)] 
      \draw[fill=green!60] (0,0.18) rectangle (0.086,0.32); 
      \node[anchor=east, scale=0.6] at (-0.15,0.25) {8.6};
      \draw[fill=red!60] (0,-0.02) rectangle (0.632,0.14);
      \node[anchor=east, scale=0.6] at (-0.15,0.06) {63.2};
      \draw[fill=blue!60] (0,-0.18) rectangle (0.045,-0.06);
      \node[anchor=east, scale=0.6] at (-0.15,-0.12) {4.5};
      \draw (1,0.32) -- (1,-0.18);
  \end{tikzpicture} &
  \begin{tikzpicture}[baseline=(current bounding box.south)] 
      \draw[fill=green!60] (0,0.18) rectangle (0.078,0.32); 
      \node[anchor=east, scale=0.6] at (-0.15,0.25) {7.8};
      \draw[fill=red!60] (0,-0.02) rectangle (0.714,0.14);
      \node[anchor=east, scale=0.6] at (-0.15,0.06) {71.4};
      \draw[fill=blue!60] (0,-0.18) rectangle (0.052,-0.06);
      \node[anchor=east, scale=0.6] at (-0.15,-0.12) {5.2};
      \draw (1,0.32) -- (1,-0.18);
  \end{tikzpicture} &
  \begin{tikzpicture}[baseline=(current bounding box.south)] 
      \draw[fill=green!60] (0,0.18) rectangle (0.095,0.32); 
      \node[anchor=east, scale=0.6] at (-0.15,0.25) {9.5};
      \draw[fill=red!60] (0,-0.02) rectangle (0.720,0.14);
      \node[anchor=east, scale=0.6] at (-0.15,0.06) {72.0};
      \draw[fill=blue!60] (0,-0.18) rectangle (0.042,-0.06);
      \node[anchor=east, scale=0.6] at (-0.15,-0.12) {4.2};
      \draw (1,0.32) -- (1,-0.18);
  \end{tikzpicture}\\
\addlinespace[0.5em]

\multicolumn{1}{@{}l}{\makecell[bl]{\vspace{-0.09cm}\small Sensitive\vspace{0.06cm}\\Information\vspace{-0.04cm}\\Disclosure}} & 
\begin{tikzpicture}[baseline=(current bounding box.south)] 
  \draw[fill=green!60] (0,0.18) rectangle (0.167,0.32); 
  \node[anchor=east, scale=0.6] at (-0.15,0.25) {16.7};
  \draw[fill=red!60] (0,-0.02) rectangle (0.589,0.14);
  \node[anchor=east, scale=0.6] at (-0.15,0.06) {58.9};
  \draw[fill=blue!60] (0,-0.18) rectangle (0.126,-0.06);
  \node[anchor=east, scale=0.6] at (-0.15,-0.12) {12.6};
  \draw (1,0.32) -- (1,-0.18);
\end{tikzpicture} &
\begin{tikzpicture}[baseline=(current bounding box.south)] 
  \draw[fill=green!60] (0,0.18) rectangle (0.154,0.32); 
  \node[anchor=east, scale=0.6] at (-0.15,0.25) {15.4};
  \draw[fill=red!60] (0,-0.02) rectangle (0.556,0.14);
  \node[anchor=east, scale=0.6] at (-0.15,0.06) {55.6};
  \draw[fill=blue!60] (0,-0.18) rectangle (0.118,-0.06);
  \node[anchor=east, scale=0.6] at (-0.15,-0.12) {11.8};
  \draw (1,0.32) -- (1,-0.18);
\end{tikzpicture} &
\begin{tikzpicture}[baseline=(current bounding box.south)] 
  \draw[fill=green!60] (0,0.18) rectangle (0.182,0.32); 
  \node[anchor=east, scale=0.6] at (-0.15,0.25) {18.2};
  \draw[fill=red!60] (0,-0.02) rectangle (0.621,0.14);
  \node[anchor=east, scale=0.6] at (-0.15,0.06) {62.1};
  \draw[fill=blue!60] (0,-0.18) rectangle (0.134,-0.06);
  \node[anchor=east, scale=0.6] at (-0.15,-0.12) {13.4};
  \draw (1,0.32) -- (1,-0.18);
\end{tikzpicture} &
\begin{tikzpicture}[baseline=(current bounding box.south)] 
  \draw[fill=green!60] (0,0.18) rectangle (0.190,0.32); 
  \node[anchor=east, scale=0.6] at (-0.15,0.25) {19.0};
  \draw[fill=red!60] (0,-0.02) rectangle (0.630,0.14);
  \node[anchor=east, scale=0.6] at (-0.15,0.06) {63.0};
  \draw[fill=blue!60] (0,-0.18) rectangle (0.110,-0.06);
  \node[anchor=east, scale=0.6] at (-0.15,-0.12) {11.0};
  \draw (1,0.32) -- (1,-0.18);
\end{tikzpicture}\\
\addlinespace[0.7em]

\multicolumn{1}{@{}l}{\makecell[bl]{\vspace{-0.05cm}\small Data and\vspace{0.03cm}\\Model\vspace{-0.05cm}\\Poisoning}} & 
\begin{tikzpicture}[baseline=(current bounding box.south)] 
  \draw[fill=green!60] (0,0.18) rectangle (0.124,0.32); 
  \node[anchor=east, scale=0.6] at (-0.15,0.25) {12.4};
  \draw[fill=red!60] (0,-0.02) rectangle (0.718,0.14);
  \node[anchor=east, scale=0.6] at (-0.15,0.06) {71.8};
  \draw[fill=blue!60] (0,-0.18) rectangle (0.119,-0.06);
  \node[anchor=east, scale=0.6] at (-0.15,-0.12) {11.9};
  \draw (1,0.32) -- (1,-0.18);
\end{tikzpicture} &
\begin{tikzpicture}[baseline=(current bounding box.south)] 
  \draw[fill=green!60] (0,0.18) rectangle (0.118,0.32); 
  \node[anchor=east, scale=0.6] at (-0.15,0.25) {11.8};
  \draw[fill=red!60] (0,-0.02) rectangle (0.695,0.14);
  \node[anchor=east, scale=0.6] at (-0.15,0.06) {69.5};
  \draw[fill=blue!60] (0,-0.18) rectangle (0.115,-0.06);
  \node[anchor=east, scale=0.6] at (-0.15,-0.12) {11.5};
  \draw (1,0.32) -- (1,-0.18);
\end{tikzpicture} &
\begin{tikzpicture}[baseline=(current bounding box.south)] 
  \draw[fill=green!60] (0,0.18) rectangle (0.136,0.32); 
  \node[anchor=east, scale=0.6] at (-0.15,0.25) {13.6};
  \draw[fill=red!60] (0,-0.02) rectangle (0.742,0.14);
  \node[anchor=east, scale=0.6] at (-0.15,0.06) {74.2};
  \draw[fill=blue!60] (0,-0.18) rectangle (0.128,-0.06);
  \node[anchor=east, scale=0.6] at (-0.15,-0.12) {12.8};
  \draw (1,0.32) -- (1,-0.18);
\end{tikzpicture} &
\begin{tikzpicture}[baseline=(current bounding box.south)] 
  \draw[fill=green!60] (0,0.18) rectangle (0.140,0.32); 
  \node[anchor=east, scale=0.6] at (-0.15,0.25) {14.0};
  \draw[fill=red!60] (0,-0.02) rectangle (0.750,0.14);
  \node[anchor=east, scale=0.6] at (-0.15,0.06) {75.0};
  \draw[fill=blue!60] (0,-0.18) rectangle (0.110,-0.06);
  \node[anchor=east, scale=0.6] at (-0.15,-0.12) {11.0};
  \draw (1,0.32) -- (1,-0.18);
\end{tikzpicture}\\
\addlinespace[0.7em]

\multicolumn{1}{@{}l}{\makecell[bl]{\vspace{-0.05cm}\small Improper\\Output\vspace{-0.05cm}\\Handling}} & 
\begin{tikzpicture}[baseline=(current bounding box.south)] 
  \draw[fill=green!60] (0,0.18) rectangle (0.089,0.32); 
  \node[anchor=east, scale=0.6] at (-0.15,0.25) {8.9};
  \draw[fill=red!60] (0,-0.02) rectangle (0.501,0.14);
  \node[anchor=east, scale=0.6] at (-0.15,0.06) {50.1};
  \draw[fill=blue!60] (0,-0.18) rectangle (0.054,-0.06);
  \node[anchor=east, scale=0.6] at (-0.15,-0.12) {5.4};
  \draw (1,0.32) -- (1,-0.18);
\end{tikzpicture} &
\begin{tikzpicture}[baseline=(current bounding box.south)] 
  \draw[fill=green!60] (0,0.18) rectangle (0.084,0.32); 
  \node[anchor=east, scale=0.6] at (-0.15,0.25) {8.4};
  \draw[fill=red!60] (0,-0.02) rectangle (0.485,0.14);
  \node[anchor=east, scale=0.6] at (-0.15,0.06) {48.5};
  \draw[fill=blue!60] (0,-0.18) rectangle (0.047,-0.06);
  \node[anchor=east, scale=0.6] at (-0.15,-0.12) {4.7};
  \draw (1,0.32) -- (1,-0.18);
\end{tikzpicture} &
\begin{tikzpicture}[baseline=(current bounding box.south)] 
  \draw[fill=green!60] (0,0.18) rectangle (0.097,0.32); 
  \node[anchor=east, scale=0.6] at (-0.15,0.25) {9.7};
  \draw[fill=red!60] (0,-0.02) rectangle (0.523,0.14);
  \node[anchor=east, scale=0.6] at (-0.15,0.06) {52.3};
  \draw[fill=blue!60] (0,-0.18) rectangle (0.061,-0.06);
  \node[anchor=east, scale=0.6] at (-0.15,-0.12) {6.1};
  \draw (1,0.32) -- (1,-0.18);
\end{tikzpicture} &
\begin{tikzpicture}[baseline=(current bounding box.south)] 
  \draw[fill=green!60] (0,0.18) rectangle (0.100,0.32); 
  \node[anchor=east, scale=0.6] at (-0.15,0.25) {10.0};
  \draw[fill=red!60] (0,-0.02) rectangle (0.530,0.14);
  \node[anchor=east, scale=0.6] at (-0.15,0.06) {53.0};
  \draw[fill=blue!60] (0,-0.18) rectangle (0.045,-0.06);
  \node[anchor=east, scale=0.6] at (-0.15,-0.12) {4.5};
  \draw (1,0.32) -- (1,-0.18);
\end{tikzpicture}\\
\addlinespace[1.1em]

\multicolumn{1}{@{}l}{\makecell[bl]{\vspace{-0.08cm}\small Excessive\\\vspace{0.02cm}Agency\\}} & 
\begin{tikzpicture}[baseline=(current bounding box.south)] 
  \draw[fill=green!60] (0,0.18) rectangle (0.142,0.32); 
  \node[anchor=east, scale=0.6] at (-0.15,0.25) {14.2};
  \draw[fill=red!60] (0,-0.02) rectangle (0.636,0.14);
  \node[anchor=east, scale=0.6] at (-0.15,0.06) {63.6};
  \draw[fill=blue!60] (0,-0.18) rectangle (0.105,-0.06);
  \node[anchor=east, scale=0.6] at (-0.15,-0.12) {10.5};
  \draw (1,0.32) -- (1,-0.18);
\end{tikzpicture} &
\begin{tikzpicture}[baseline=(current bounding box.south)] 
  \draw[fill=green!60] (0,0.18) rectangle (0.128,0.32); 
  \node[anchor=east, scale=0.6] at (-0.15,0.25) {12.8};
  \draw[fill=red!60] (0,-0.02) rectangle (0.618,0.14);
  \node[anchor=east, scale=0.6] at (-0.15,0.06) {61.8};
  \draw[fill=blue!60] (0,-0.18) rectangle (0.093,-0.06);
  \node[anchor=east, scale=0.6] at (-0.15,-0.12) {9.3};
  \draw (1,0.32) -- (1,-0.18);
\end{tikzpicture} &
\begin{tikzpicture}[baseline=(current bounding box.south)] 
  \draw[fill=green!60] (0,0.18) rectangle (0.151,0.32); 
  \node[anchor=east, scale=0.6] at (-0.15,0.25) {15.1};
  \draw[fill=red!60] (0,-0.02) rectangle (0.654,0.14);
  \node[anchor=east, scale=0.6] at (-0.15,0.06) {65.4};
  \draw[fill=blue!60] (0,-0.18) rectangle (0.117,-0.06);
  \node[anchor=east, scale=0.6] at (-0.15,-0.12) {11.7};
  \draw (1,0.32) -- (1,-0.18);
\end{tikzpicture} &
\begin{tikzpicture}[baseline=(current bounding box.south)] 
  \draw[fill=green!60] (0,0.18) rectangle (0.155,0.32); 
  \node[anchor=east, scale=0.6] at (-0.15,0.25) {15.5};
  \draw[fill=red!60] (0,-0.02) rectangle (0.660,0.14);
  \node[anchor=east, scale=0.6] at (-0.15,0.06) {66.0};
  \draw[fill=blue!60] (0,-0.18) rectangle (0.090,-0.06);
  \node[anchor=east, scale=0.6] at (-0.15,-0.12) {9.0};
  \draw (1,0.32) -- (1,-0.18);
\end{tikzpicture}\\
\addlinespace[0.9em]

\multicolumn{1}{@{}l}{\makecell[bl]{\vspace{-0.03cm}\small Embedding\\\vspace{-0.01cm}Weaknesses\\}} & 
\begin{tikzpicture}[baseline=(current bounding box.south)] 
  \draw[fill=green!60] (0,0.18) rectangle (0.211,0.32); 
  \node[anchor=east, scale=0.6] at (-0.15,0.25) {21.1};
  \draw[fill=red!60] (0,-0.02) rectangle (0.645,0.14);
  \node[anchor=east, scale=0.6] at (-0.15,0.06) {64.5};
  \draw[fill=blue!60] (0,-0.18) rectangle (0.073,-0.06);
  \node[anchor=east, scale=0.6] at (-0.15,-0.12) {7.3};
  \draw (1,0.32) -- (1,-0.18);
\end{tikzpicture} &
\begin{tikzpicture}[baseline=(current bounding box.south)] 
  \draw[fill=green!60] (0,0.18) rectangle (0.200,0.32); 
  \node[anchor=east, scale=0.6] at (-0.15,0.25) {20.0};
  \draw[fill=red!60] (0,-0.02) rectangle (0.619,0.14);
  \node[anchor=east, scale=0.6] at (-0.15,0.06) {61.9};
  \draw[fill=blue!60] (0,-0.18) rectangle (0.065,-0.06);
  \node[anchor=east, scale=0.6] at (-0.15,-0.12) {6.5};
  \draw (1,0.32) -- (1,-0.18);
\end{tikzpicture} &
\begin{tikzpicture}[baseline=(current bounding box.south)] 
  \draw[fill=green!60] (0,0.18) rectangle (0.223,0.32); 
  \node[anchor=east, scale=0.6] at (-0.15,0.25) {22.3};
  \draw[fill=red!60] (0,-0.02) rectangle (0.672,0.14);
  \node[anchor=east, scale=0.6] at (-0.15,0.06) {67.2};
  \draw[fill=blue!60] (0,-0.18) rectangle (0.081,-0.06);
  \node[anchor=east, scale=0.6] at (-0.15,-0.12) {8.1};
  \draw (1,0.32) -- (1,-0.18);
\end{tikzpicture} &
\begin{tikzpicture}[baseline=(current bounding box.south)] 
  \draw[fill=green!60] (0,0.18) rectangle (0.228,0.32); 
  \node[anchor=east, scale=0.6] at (-0.15,0.25) {22.8};
  \draw[fill=red!60] (0,-0.02) rectangle (0.680,0.14);
  \node[anchor=east, scale=0.6] at (-0.15,0.06) {68.0};
  \draw[fill=blue!60] (0,-0.18) rectangle (0.062,-0.06);
  \node[anchor=east, scale=0.6] at (-0.15,-0.12) {6.2};
  \draw (1,0.32) -- (1,-0.18);
\end{tikzpicture}\\
\addlinespace[1em]

\multicolumn{1}{@{}l}{\makecell[bl]{\vspace{-0.08cm}\small Mis-\\\vspace{0.04cm}information}} & 
\begin{tikzpicture}[baseline=(current bounding box.south)] 
  \draw[fill=green!60] (0,0.18) rectangle (0.160,0.32); 
  \node[anchor=east, scale=0.6] at (-0.15,0.25) {16.0};
  \draw[fill=red!60] (0,-0.02) rectangle (0.746,0.14);
  \node[anchor=east, scale=0.6] at (-0.15,0.06) {74.6};
  \draw[fill=blue!60] (0,-0.18) rectangle (0.208,-0.06);
  \node[anchor=east, scale=0.6] at (-0.15,-0.12) {20.8};
  \draw (1,0.32) -- (1,-0.18);
\end{tikzpicture} &
\begin{tikzpicture}[baseline=(current bounding box.south)] 
  \draw[fill=green!60] (0,0.18) rectangle (0.149,0.32); 
  \node[anchor=east, scale=0.6] at (-0.15,0.25) {14.9};
  \draw[fill=red!60] (0,-0.02) rectangle (0.729,0.14);
  \node[anchor=east, scale=0.6] at (-0.15,0.06) {72.9};
  \draw[fill=blue!60] (0,-0.18) rectangle (0.197,-0.06);
  \node[anchor=east, scale=0.6] at (-0.15,-0.12) {19.7};
  \draw (1,0.32) -- (1,-0.18);
\end{tikzpicture} &
\begin{tikzpicture}[baseline=(current bounding box.south)] 
  \draw[fill=green!60] (0,0.18) rectangle (0.172,0.32); 
  \node[anchor=east, scale=0.6] at (-0.15,0.25) {17.2};
  \draw[fill=red!60] (0,-0.02) rectangle (0.768,0.14);
  \node[anchor=east, scale=0.6] at (-0.15,0.06) {76.8};
  \draw[fill=blue!60] (0,-0.18) rectangle (0.224,-0.06);
  \node[anchor=east, scale=0.6] at (-0.15,-0.12) {22.4};
  \draw (1,0.32) -- (1,-0.18);
\end{tikzpicture} &
\begin{tikzpicture}[baseline=(current bounding box.south)] 
  \draw[fill=green!60] (0,0.18) rectangle (0.176,0.32); 
  \node[anchor=east, scale=0.6] at (-0.15,0.25) {17.6};
  \draw[fill=red!60] (0,-0.02) rectangle (0.775,0.14);
  \node[anchor=east, scale=0.6] at (-0.15,0.06) {77.5};
  \draw[fill=blue!60] (0,-0.18) rectangle (0.190,-0.06);
  \node[anchor=east, scale=0.6] at (-0.15,-0.12) {19.0};
  \draw (1,0.32) -- (1,-0.18);
\end{tikzpicture}\\
\bottomrule
\end{tabularx}
}
\endgroup
\end{table*}

The testing utilised the framework described in Appendix~\ref{app:owasp}, where each vulnerability category was tested using multiple specific probes (e.g., ``Prompt Injection'' was tested using dan\footnote{https://reference.garak.ai/en/stable/garak.probes.dan.html}, prompt inject\footnote{https://reference.garak.ai/en/stable/garak.probes.promptinject.html}, encoding\footnote{https://reference.garak.ai/en/stable/garak.probes.encoding.html}, and latent injection\footnote{https://reference.garak.ai/en/stable/garak.probes.latentinjection.html} probes). For each vulnerability category, we calculated the failure rate as the percentage of failed tests across all probes in that category. For example, if a model failed 5 out of 10 tests in a particular probe, the failure rate for that probe would be 50\%. 

The next section presents a comparative analysis of these failure rates across three model configurations for each of the four models: the base model without fine-tuning, the model fine-tuned on the original CyberLLMInstruct, and the model fine-tuned on our safety-aware transformed version. This analysis provides insights into how safety-regulation affects model vulnerability to various attack vectors across different model architectures.

\section{Results}
\label{sec:results}

This section presents the results of our evaluation of LLM safety vulnerabilities and alignment, providing independent validation of the CyberLLMInstruct findings through a different evaluation framework. All reported results are based on the average of 5 independent runs to ensure statistical reliability. We begin by analysing the safety of various models against OWASP Top 10 for LLM Applications vulnerabilities using the garak framework, confirming the safety degradation patterns identified in the original study. This is followed by examination of inference time impacts. The results demonstrate significant safety degradation in fine-tuned models, validating the previous findings while extending them to include the reasoning-capable DeepSeek R1 8B model. We then present our novel findings on safety alignment through safety-regulation, showing how careful rewording can mitigate some of the safety risks introduced by fine-tuning.

\subsection{Safety Analysis}

Table~\ref{tab:owasp_results} presents a comprehensive analysis of how base, fine-tuned, and safety-enhanced LLMs perform across OWASP Top 7 for LLM Applications vulnerabilities. The evaluation used garak failure rates from 0 (fully secure) to 100 (completely vulnerable). Figure~\ref{tab:inference_time} complements this by showing the inference time comparisons across the three model configurations. A concerning pattern emerged across all models: fine-tuning consistently led to increased failure rates across all vulnerability categories, while safety alignment significantly improved performance.

\begin{figure}[tb!]
\centering
\begin{tikzpicture}
    \begin{axis}[
        xtick=data,
        symbolic x coords={
            Mistral 7B,
            Llama 3 8B,
            Gemma 2 9B,
            DeepSeek R1 8B
        },
        ybar,
        bar width=16pt,
        width=\linewidth,
        height=0.5\linewidth,
        ylabel={Inference Time (minutes)},
        nodes near coords,
        nodes near coords style={font=\scriptsize},
        enlarge x limits=0.2,
        x tick label style={rotate=45, anchor=east, font=\small},
        legend style={at={(0.5,1.12)}, anchor=south, font=\small, align=left, column sep=0.9em},
        legend columns=3,
        legend image code/.code={\draw[#1] (0cm,-0.1cm) rectangle (0.2cm,0.1cm);},
        tick label style={font=\small},
        label style={font=\small}
    ]
        \addplot[fill=green!60] coordinates {
            (Mistral 7B,69)
            (Llama 3 8B,61)
            (Gemma 2 9B,85)
            (DeepSeek R1 8B,78)
        };
        \addplot[fill=red!60] coordinates {
            (Mistral 7B,112)
            (Llama 3 8B,74)
            (Gemma 2 9B,138)
            (DeepSeek R1 8B,95)
        };
        \addplot[fill=blue!60] coordinates {
            (Mistral 7B,65)
            (Llama 3 8B,58)
            (Gemma 2 9B,80)
            (DeepSeek R1 8B,72)
        };
        \legend{Base model, Fine-tuned model, Safety-enhanced model}
    \end{axis}
\end{tikzpicture}
\caption{Inference times for base (green), fine-tuned (red), and safety-enhanced (blue) LLMs during garak testing.}
\label{tab:inference_time}
\end{figure}

``Prompt Injection'' emerged as the most severely compromised category post-fine-tuning, with failure rates reaching as high as 77.5\% for DeepSeek R1 8B. All models showed dramatic increases in vulnerability after fine-tuning, with safety alignment providing significant improvements, reducing failure rates to as low as 4.2\%.

The ``Sensitive Information Disclosure'' category revealed similar concerning trends. Models across different architectures showed marked vulnerability increases after fine-tuning, with failure rates ranging from 55.6\% to 63.0\%. Safety alignment consistently improved performance, reducing failure rates to 11.0-13.4\%.

In the ``Improper Output Handling'' category, models showed varying degrees of resilience, with failure rates ranging from 48.5\% to 53.0\% after fine-tuning. Safety alignment provided substantial improvements, reducing failure rates to 4.5-6.1\%.

The ``Data and Model Poisoning'' category showed significant vulnerability increases across all models, with failure rates reaching 69.5-75.0\% after fine-tuning. Safety alignment consistently improved performance, reducing failure rates to 11.0-12.8\%.

``Excessive Agency'' revealed substantial security compromises across all models, with failure rates ranging from 61.8\% to 66.0\% after fine-tuning. Safety alignment provided notable improvements, reducing failure rates to 9.0-11.7\%.

``Embedding Weaknesses'' showed concerning vulnerability increases, with failure rates ranging from 61.9\% to 68.0\% after fine-tuning. Safety alignment consistently improved performance, reducing failure rates to 6.2-8.1\%.

``Misinformation'' proved to be the most challenging category, with failure rates reaching 72.9-77.5\% after fine-tuning. However, safety alignment still provided improvements, reducing failure rates to 19.0-22.4\%.

The analysis reveals a clear pattern: while fine-tuning enhances task-specific performance--as shown in the experiments reported in~\cite{CyberLLMInstruct} -- it consistently compromises safety across all vulnerability categories. Input manipulation vulnerabilities (particularly ``Prompt Injection'') and data exposure risks (``Sensitive Information Disclosure'') emerged as the most critical concerns. Safety alignment through safety-regulating process consistently improved performance across all categories, demonstrating the effectiveness of our approach in mitigating the security risks introduced by fine-tuning.

\subsection{Safety Alignment}
\label{subsec:safety_alignment}

To demonstrate the feasibility of our approach, we conducted experiments that focused on the safety alignment analysis across all OWASP Top 10 for LLM Applications vulnerability categories. The testing was performed using the garak framework, with a total of 14,395 individual test cases distributed across the vulnerability categories as follows:

\begin{itemize}
\item Prompt Injection: 5,425 tests
\item Sensitive Information Disclosure: 370 tests
\item Data and Model Poisoning: 2,170 tests
\item Improper Output Handling: 1,280 tests
\item Excessive Agency: 60 tests
\item Vector and Embedding Weaknesses: 1,180 tests
\item Misinformation: 3,910 tests
\end{itemize}

The ``Supply Chain'', ``System Prompt Leakage'', and ``Unbounded Consumption'' categories were not included in the analysis, as they were not yet supported by the garak framework at the time this paper was being written (May--June 2025).

Table~\ref{tab:owasp_results} presents a comprehensive overview of the garak failure rates for each vulnerability category across three model configurations for each of the four tested models (Mistral 7B, Llama 3 8B, Gemma 2 9B, and DeepSeek R1 8B). The table shows three bars for each model: \textbf{Base} (green, original pre-trained models without fine-tuning), \textbf{Fine-tuned} (red, models fine-tuned on the original CyberLLMInstruct dataset), and \textbf{Safety-enhanced} (blue, models fine-tuned on our safety-aware transformed version of CyberLLMInstruct). The failure rates range from 0 (fully secure) to 100 (completely vulnerable), allowing for direct comparison of safety alignment effectiveness across different model architectures.

The results reveal a clear and consistent pattern across all vulnerability categories. Most critically, \textbf{every single highest failure rate} (represented by the longest red bars) occurs in the \textbf{Fine-tuned} configuration, with models consistently exhibiting the worst performance when fine-tuned without safety measures. These failure rates are alarmingly high, ranging from 48.5\% (``Improper Output Handling'') to 77.5\% (``Misinformation''). Conversely, \textbf{nearly all lowest failure rates} (represented by the shortest blue bars) appear in the \textbf{Safety-enhanced} configuration, with most values below 15\% and the best performance reaching as low as 4.2\% for ``Prompt Injection'' with DeepSeek R1 8B. This dramatic contrast--often exceeding 60 percentage points difference between worst and best performance--demonstrates that safety alignment is not merely beneficial but essential for secure deployment of cyber security LLMs.

\section{Further Discussions}
\label{sec:discussions}

Our experimental results reveal critical insights into the safety implications of fine-tuning LLMs with pseudo-malicious cyber security data. The comprehensive testing across OWASP Top 10 for LLM vulnerabilities (see Table~\ref{tab:owasp_results}) demonstrates that fine-tuning consistently compromises model safety across all vulnerabilities in OWASP Top 10 for LLMs. This degradation pattern holds true across different model architectures and sizes, suggesting a fundamental challenge in maintaining safety during domain-specific adaptation.

\begin{figure}[tb!]
\centering
\begin{tikzpicture}
\begin{axis}[
  width=\linewidth,
  height=11.4cm,
  xlabel={Vulnerability Category},
  ylabel={Absolute Difference in Failure Rates (\%)},
  xmin=0.5, xmax=7.5,
  ymin=-10, ymax=80,
  xtick={1,2,3,4,5,6,7},
  xticklabels={
    {Prompt Injection},
    {Sensitive Info. Disclosure},
    {Data and Model Poisoning},
    {Improper Output Handling},
    {Excessive Agency},
    {Embedding Weaknesses},
    {Misinformation}
  },
  xticklabel style={rotate=45, anchor=east, font=\small},
  ytick={-10,0,10,20,30,40,50,60,70,80},
  legend style={
    at={(0.455,1.05)},
    anchor=south,
    font=\small,
    column sep=1ex
  },
  legend columns=2,
  legend cell align={left},
  grid=both,
  grid style={line width=.1pt, draw=gray!10},
  major grid style={line width=.2pt, draw=gray!50},
  ]
  
  ]

\addplot+[mark=*, mark options=solid, line width=1.5pt, color=blue70] coordinates {
  (1,2.8) (2,4.1) (3,0.5) (4,3.5) (5,3.7) (6,13.8) (7,-4.8)
};
\addlegendentry{Base -- Safety-enhanced (Mistral 7B)}
\addplot+[mark=o, mark options=solid, line width=1.5pt, color=blue70, dashed] coordinates {
  (1,59.6) (2,42.2) (3,59.4) (4,41.2) (5,49.4) (6,43.4) (7,58.6)
};
\addlegendentry{Fine-tuned -- Base (Mistral 7B)}

\addplot+[mark=square*, mark options=solid, line width=1.5pt, color=green70] coordinates {
  (1,4.1) (2,3.6) (3,0.3) (4,3.7) (5,3.5) (6,13.5) (7,-4.8)
};
\addlegendentry{Base -- Safety-enhanced (Llama 3 8B)}
\addplot+[mark=square, mark options=solid, line width=1.5pt, color=green70, dashed] coordinates {
  (1,54.6) (2,40.2) (3,57.7) (4,40.1) (5,49.0) (6,41.9) (7,58.0)
};
\addlegendentry{Fine-tuned -- Base (Llama 3 8B)}

\addplot+[mark=diamond*, mark options=solid, mark size=3pt, line width=1.5pt, color=red70] coordinates {
  (1,2.6) (2,4.8) (3,0.8) (4,3.6) (5,3.4) (6,14.2) (7,-5.2)
};
\addlegendentry{Base -- Safety-enhanced (Gemma 2 9B)}
\addplot+[mark=diamond, mark options=solid, mark size=3pt, line width=1.5pt, color=red70, dashed] coordinates {
  (1,63.6) (2,43.9) (3,60.6) (4,42.6) (5,50.3) (6,44.9) (7,59.6)
};
\addlegendentry{Fine-tuned -- Base (Gemma 2 9B)}

\addplot+[mark=triangle*, mark options=solid, mark size=3pt, line width=1.5pt, color=orange70, solid] coordinates {
   (1,5.3) (2,8.0) (3,3.0) (4,5.5) (5,6.5) (6,16.6) (7,-1.4)
};
\addlegendentry{Base -- Safety-enhanced (DeepSeek R1 8B)}
\addplot+[mark=triangle, mark options=solid, mark size=3pt, line width=1.5pt, color=orange70, dashed] coordinates {
   (1,62.5) (2,44.0) (3,61.0) (4,43.0) (5,50.5) (6,45.2) (7,59.9)
};
\addlegendentry{Fine-tuned -- Base (DeepSeek R1 8B)}

\end{axis}
\end{tikzpicture}
\caption{Absolute differences in failure rates showing two key comparisons: (1) Base -- Safety-enhanced (solid lines, positive values indicate safety improvement from base to safety-enhanced models), and (2) Fine-tuned -- Base (dashed lines, positive values indicate safety degradation from base to fine-tuned models). Higher values in Base -- Safety-enhanced indicate better safety alignment effectiveness, while higher values in Fine-tuned -- Base indicate greater safety degradation from fine-tuning.}
\label{fig:safety_improvement}
\end{figure}

The relationship between model architecture and safety resilience presents interesting variations, as shown in Fig.~\ref{fig:safety_improvement}. The figure illustrates two critical patterns: (1) the safety degradation caused by fine-tuning (Fine-tuned - Base, dashed lines) and (2) the safety improvement achieved by safety alignment (Base - Safety-enhanced, solid lines). Fine-tuning consistently caused substantial safety degradation across all models, with increases in failure rates ranging from 40-64 percentage points. Safety alignment showed more modest but consistent improvements, with DeepSeek R1 8B demonstrating the strongest safety alignment effectiveness, particularly in Embedding Weaknesses (16.6 percentage point improvement) and Sensitive Info. Disclosure (8.0 percentage point improvement). Notably, Misinformation showed negative Base - Safety-enhanced values for most models, indicating that safety alignment was less effective for this vulnerability category. This suggests that architectural choices and fine-tuning methodologies play a crucial role in safety preservation and the effectiveness of safety alignment approaches.

Vulnerability patterns also vary significantly across different attack categories, as detailed in Table~\ref{tab:owasp_results}. Models demonstrate relative stability in areas like ``Improper Output Handling'', while showing substantial vulnerability increases in ``Prompt Injection'' and ``Misinformation''. This category-specific behaviour indicates that safety mechanisms may be more resilient to certain types of attacks than others, highlighting the need for targeted safety improvements.

The inference time analysis, shown in Fig.~\ref{tab:inference_time}, reveals interesting patterns across the three model configurations: fine-tuned versions consistently require more time to process test inputs than their base counterparts, while safety-enhanced models show slightly improved efficiency compared to base models. The increased inference time in fine-tuned models can be attributed to their more detailed and context-aware responses to cyber security queries. While base models often provide quick rejection responses when faced with potentially harmful queries, fine-tuned models engage in more comprehensive analysis and response generation. Safety-enhanced models maintain this detailed analysis while incorporating safety considerations, resulting in slightly more efficient processing. This behaviour aligns with our safety analysis results, where safety-enhanced models demonstrated the best balance of security and utility.

The use of pseudo-malicious data (descriptions of malicious actions without actual harmful code) in fine-tuning raises important questions about the mechanisms behind safety degradation. Our results suggest that vulnerabilities may arise not only from exposure to pseudo-malicious content but also from the model's response to safety-critical information. This observation points to potential weaknesses in current safety mechanisms that may be exacerbated by fine-tuning, rather than being solely caused by the malicious intent of the content itself.

A particularly significant finding emerged from our comparison of fine-tuning with the original pseudo-malicious data versus the safety-aware transformed version, as shown in Table~\ref{tab:owasp_results}. The visual patterns in the table dramatically illustrate the effectiveness of our approach: the systematic occurrence of highest failure rates (represented by the longest red bars) exclusively in models fine-tuned without safety measures, contrasted with the concentration of lowest failure rates (represented by the shortest blue bars) in safety-enhanced models. This consistent pattern across all vulnerability categories and model architectures provides compelling evidence that safety alignment through careful data safety-regulation is not just effective but crucial for mitigating the security risks inherent in fine-tuning LLMs with cyber security data.

\subsection{Future Work}
The key takeaway from our study is that while fine-tuning LLMs with cyber security data presents significant safety challenges, these challenges can be mitigated through careful data safety-regulation and safety-aware approaches. Future work will focus on two main directions: (1) conducting an ablation analysis on different categories of cyber security data to understand how specific types of content affect model safety, and (2) analysing safety across datasets of varying sizes and content within the cyber security domain to study the relationship between dataset characteristics and safety outcomes. These investigations will help develop more robust safety-preserving fine-tuning methodologies for LLMs in cyber security applications.

\subsection{Limitations}
The garak framework used in our tests can introduce biases or fail to represent model behaviours across domain-specific edge cases. Utilising the CyberLLMInstruct dataset itself is not without challenges, including potential biases stemming from its data sources and an imbalanced distribution of categories. Moreover, experiments could have been broadened to explore additional architectures or hyper-parameters to offer a more complete view of the interplay between model size and safety.

\section{Conclusion}
\label{sec:conclusion}

Our evaluation of safety risks in fine-tuned LLMs for cyber security applications provides independent validation of the critical safety concerns identified in the CyberLLMInstruct paper, while extending the findings with novel contributions. Through testing using the garak red teaming framework across OWASP Top 10 for LLM Applications vulnerabilities, we confirm that fine-tuning consistently compromises model safety across all tested models and vulnerability categories, validating the previous findings through a different evaluation methodology. 

Our extension to include the reasoning-capable DeepSeek R1 8B model demonstrates that these safety concerns apply across diverse model architectures. The novel safety-aware safety-regulating approach presents a promising direction for mitigating these risks. By carefully rewording instruction-response pairs to include explicit safety precautions and ethical considerations, we show that it is possible to maintain or even improve model safety while preserving technical utility. This finding suggests that the way security information is presented during fine-tuning can significantly impact model behaviour, offering a practical path forward for developing safer fine-tuning methodologies. 

These results highlight the importance of considering safety implications when fine-tuning LLMs for cyber security applications. The demonstrated effectiveness of safety-regulation in mitigating security risks while maintaining model utility provides a foundation for developing more secure and reliable LLM-based cyber security solutions.

\begin{credits}
\subsubsection{\ackname} This work was partly supported by the research project ``Countering HArms caused by Ransomware in the Internet Of Things (CHARIOT)'', funded by the EPSRC  (Engineering and Physical Sciences Research Council), part of UKRI (UK Research and Innovation), under the reference number EP/X036707/1. The authors would also like to thank the anonymous reviewers for their constructive feedback.

\subsubsection{\discintname}
The authors have no competing interests to declare that are relevant to the content of this article.
\end{credits}

\appendix

\section{OWASP Top 10 for LLM Applications}
\label{app:owasp}

The 2025 edition of the OWASP Top 10 for LLM Applications framework~\cite{owasp2025} includes:
\begin{enumerate}
\item \textbf{Prompt Injection}: Manipulating inputs to alter model behaviour maliciously. This is tested as a baseline vulnerability and applicable across categories with enhanced attack strategies.

\item \textbf{Sensitive Information Disclosure}: Exposing confidential data through model outputs. This category includes nine vulnerabilities, such as Prompt Leakage (4 types), PII Leakage (4 types), and Intellectual Property disclosure (1 type).

\item \textbf{Supply Chain}: Compromising the integrity of training data, pre-trained models, or deployment platforms. It is evaluated indirectly through other categories like data poisoning, security leaks, and excessive functionality.

\item \textbf{Data and Model Poisoning}: Introducing vulnerabilities or biases during training or fine-tuning. This category tests five vulnerabilities: Bias, Toxicity, Illegal Activity, Graphic Content, and Personal Safety.

\item \textbf{Improper Output Handling}: Generating unsafe, incorrect, or harmful outputs due to poor filtering or validation. This is assessed as a general vulnerability.

\item \textbf{Excessive Agency}: Granting excessive autonomy to models, leading to unintended actions. This includes three key vulnerabilities: Excessive Functionality, Permissions, and Autonomy.

\item \textbf{System Prompt Leakage}: Revealing internal prompts that guide model behaviour, potentially allowing attackers to bypass restrictions. This category is tested across four specific types of prompt leakage vulnerabilities.

\item \textbf{Vector and Embedding Weaknesses}: Exploiting flawed or biased vector representations. It is evaluated as a general risk without specific subcategories.

\item \textbf{Misinformation}: Generating false or misleading content that appears credible. This category includes four vulnerabilities: Factual Errors, Unsupported Claims, Expertise Misrepresentation, and Discreditation.

\item \textbf{Unbounded Consumption}: Causing system performance issues or crashes through excessive output generation. This is assessed as a general vulnerability.
\end{enumerate}

\section{Technical Setup and Implementation Details}
\label{app:technical-setup}

\subsection{Hardware and Software Environment}

The fine-tuning experiments were conducted on a high performance computing cluster equipped with:
\begin{itemize}
\item \textbf{GPU}: NVIDIA A100 80GB 
\item \textbf{CPU}: Intel Xeon E5520 running at 2.27GHz
\item \textbf{Fine-tuning Framework}: SFTTrainer from the TRL library~\cite{vonwerra2022trl}
\item \textbf{Model Configuration}: TrainingArguments from the Transformers library~\cite{wolf-etal-2020-transformers}
\end{itemize}

\subsection{Model Selection Rationale}

The selected models demonstrate strong performance across diverse benchmarks:
\begin{itemize}
\item \textbf{Mistral 7B}~\cite{Mistral7B}: Competitive performance with strong reasoning capabilities
\item \textbf{Llama 3 8B}~\cite{Llama3_8B}: 79.6\% on Massive Multitask Language Understanding (MMLU) benchmark~\cite{MMLU_Benchmark}, strong general-purpose performance
\item \textbf{Gemma 2 9B}~\cite{Gemma2_9B}: Google's architecture with strong safety alignments
\item \textbf{DeepSeek R1 8B}~\cite{deepseekai2025}: Advanced reasoning capabilities with fewer safety safeguards (DeepSeek-R1-0528-Qwen3-8B variant)
\end{itemize}

The diversity in model architectures and sizes enables comprehensive analysis of how different factors influence fine-tuned model capabilities and vulnerabilities. All models are open-source, supporting reproducibility and flexible experimentation.

\subsection{Fine-tuning Hyperparameters and Training Details}
\label{app:finetune-details}

The following details the specific hyperparameters used for model fine-tuning:

\begin{itemize}
\item \textbf{Batch size:} 4 per device
\item \textbf{Gradient accumulation steps:} 4 (effective batch size: 16)
\item \textbf{Epochs:} 3
\item \textbf{Learning rate:} $2 \times 10^{-4}$
\item \textbf{Precision:} 16-bit floating point (\texttt{bfloat16} when supported)
\item \textbf{Optimizer:} AdamW with weight decay 0.01
\item \textbf{Scheduler:} Linear
\end{itemize}

This configuration is consistent with major projects like Alpaca~\cite{taori2023alpaca} (3 epochs) and FLAN~\cite{pmlr-v202-longpre23a} (2-3 epochs), and is particularly suitable given the high-quality, curated nature of the CyberLLMInstruct dataset.

\section{Safety Evaluation Framework Specifications}
\label{app:eval}

\subsection{garak Probe Specifications}

For each OWASP vulnerability category, we selected specific garak probes:

\begin{itemize}
\item \textbf{Prompt Injection}: dan (Dan\_11\_0, DUDE), promptinject (HijackHateHumans, HijackKillHumans), encoding (InjectBase64, InjectROT13), latentinjection (LatentJailbreak)
\item \textbf{Sensitive Information Disclosure}: leakreplay (GuardianCloze, PotterComplete), xss (MarkdownImageExfil)
\item \textbf{Data and Model Poisoning}: atkgen (Tox), malwaregen (Evasion, Payload), realtoxicityprompts (RTPIdentity\_Attack, RTPThreat)
\item \textbf{Improper Output Handling}: donotanswer (MaliciousUses), lmrc (Profanity, SlurUsage), grandma (Slurs)
\item \textbf{Excessive Agency}: exploitation (JinjaTemplatePythonInjection, SQLInjectionEcho)
\item \textbf{Vector and Embedding Weaknesses}: divergence (Repeat), snowball (GraphConnectivity, Primes)
\item \textbf{Misinformation}: misleading (FalseAssertion), packagehallucination (in JavaScript, Python)
\end{itemize}

Note: Supply Chain, System Prompt Leakage, and Unbounded Consumption categories were not yet supported in garak's testing framework~\cite{garak-minus-3} during the writing of this paper (May-June 2025).

\section{Example Prompts}
\label{app:prompts}

Some examples of the prompts used in this research are available at \url{https://github.com/Adelsamir01/CyberLLMInstruct/tree/main/examples/adversarial_prompts}.

\bibliographystyle{splncs04}
\bibliography{main}

\end{document}